\documentclass[article]{aa} 

\usepackage{graphicx}
\usepackage{multirow}
\usepackage[style=base]{subcaption}
\usepackage{txfonts}
\usepackage[colorlinks = true, urlcolor = blue, citecolor = black]{hyperref}
\begin{document}

   \title{\textbf{A compact group lens modeled with GIGA-Lens: Enhanced inference for complex systems}}

   \author{F.~Urcelay \inst{1}
           \and
           E.~Jullo \inst{2}
           \and
           L.~F.~Barrientos \inst{1}
           \and
           X.~Huang \inst{3} 
           \and
           J.~Hernandez \inst{1}
          }

   \institute{%
   Institute of Astrophysics, Pontificia Universidad Cat\'olica de Chile, Santiago, Chile \\
   e-mail: fjurcelay@uc.cl
   \and
    Aix Marseille Univ, CNRS, CNES, LAM, Marseille, France
    \and
    Department of Physics \& Astronomy, University of San Francisco, San Francisco, USA
    }

    \date{Received Month day, year; accepted Month day, year}
 
    \abstract
    {In the era of large-scale astronomical surveys, fast modeling of strong lens systems has become increasingly vital. While significant progress has been made for galaxy-scale lenses, the development of automated methods for modeling larger systems, such as groups and clusters, is not as extensive.}
    {Our study aims to extend the capabilities of the GIGA-Lens code, enhancing its efficiency in modeling multi-galaxy strong lens systems. We focus on demonstrating the potential of GPU-accelerated Bayesian inference in handling complex lensing scenarios with a high number of free parameters.}
    {
    We employ an improved inference approach that combines image position and pixelated data with an annealing sampling technique to obtain the posterior distribution of complex models. 
    This method allows us to overcome the challenge of limited prior information, a high number of parameters, and memory usage. Our process is exemplified through the analysis of the compact group lens system \object{DES J0248-3955}, for which we present VLT/X-shooter spectra.}
    {We measure a redshift of $z = 0.69 \pm 0.04$ for the group, and $z = 1.2722 \pm 0.0005$ for one of the extended arcs. Our enhanced method successfully constrained a lens model with 29 free parameters and lax priors in a remarkably short time. The mass of the lens is well described by a single dark-matter halo with a velocity dispersion of $\sigma_v = (690 \pm 30) \, km \, s^{-1}$. The model predicts the presence of a second source at the same redshift and a third source at approximately $z \sim 2.7$.}
    {Our study demonstrates the effectiveness of our lens modeling technique for dealing with a complex system in a short time using ground-based data. 
    This presents considerable potential within the context of large surveys such as LSST.
    }
    

    \maketitle
%


\begin{figure*}[t]
    \begin{subfigure}[b]{\linewidth}
    \centering
     \includegraphics[width=1\textwidth]{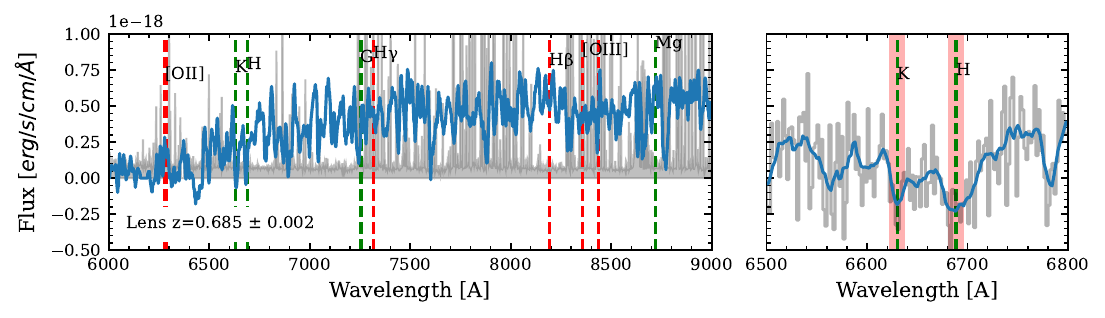}
    \label{fig:lens_spec}
    \end{subfigure}
    \begin{subfigure}[b]{\linewidth}
    \centering
    \includegraphics[width=1\textwidth]{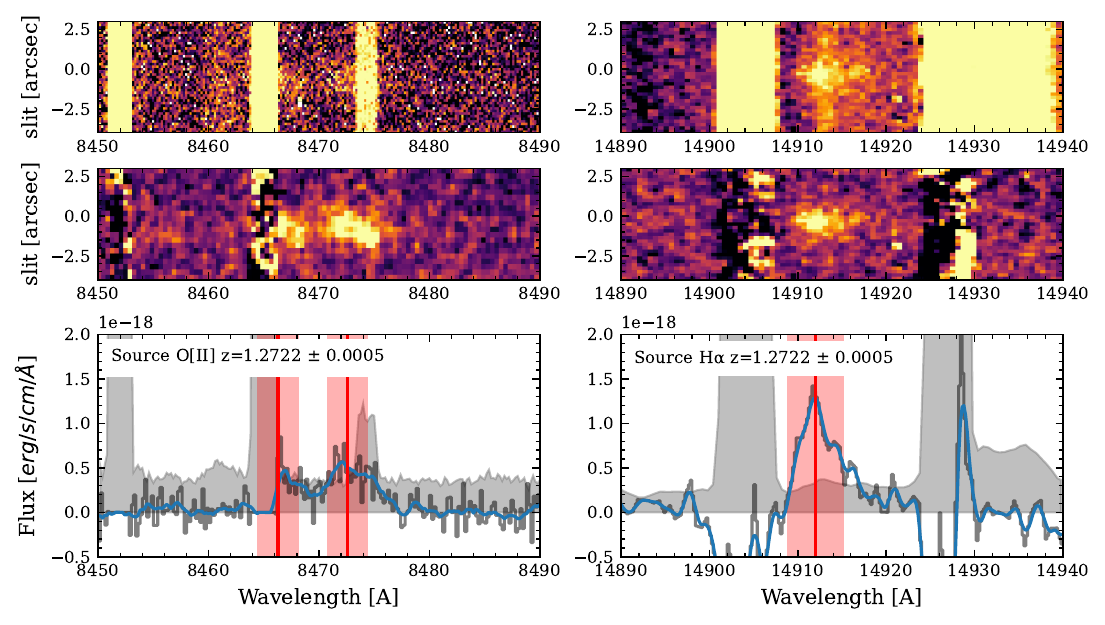}
    \label{fig:source_spec}
    \end{subfigure}
    \caption{%
    \textbf{Top panels:} X-Shooter 1D spectrum of the BGG at redshift $z=0.685 \pm 0.002$. The redshift is driven by the $K$ and $H$ absorption lines (right panel), as well as the shape of the continuum. The red shaded region around the $K$ and $H$ lines represents the associated uncertainty. 
    \textbf{Second-row panels:} 2D spectrum of the source S1a.1 at redshift $z=1.2722 \pm 0.0005$ without sky subtraction, centered around the [O II] (left) and $H_\alpha$ (right).
    \textbf{Third row panels:} 2D spectrum with sky subtraction applied. The [O II] spectrum has been rebinned by a factor of 2x2 to enhance the signal-to-noise ratio.
    \textbf{Bottom panels:} 1D sky-subtracted spectra of both [O II] and $H_\alpha$ emission lines at redshift $z=1.2722 \pm 0.0005$, the red line shows the expected position of the line and the red shaded region shows the uncertainty.
    The red vertical line marks the expected position of each line, and the red shaded region represents the associated uncertainty. In all 1D spectra, the gray histogram displays the unbinned spectrum, the blue curve shows the same spectrum with a binning of 3 pixels, and the gray shaded region represents the rescaled sky level for visualization purposes.
    }
    \label{fig:spectrum}
\end{figure*}

\section{Introduction} \label{sec:intro}
Strong gravitational lenses are a valuable tool in astrophysics with diverse applications. They enable us to measure the mass distribution of galaxies and clusters
\citep{Mellier1993, Limousin2005, Auger2010, Sharon2020, Limousin2022},
sometimes providing enough sensitivity to constrain the nature of dark matter 
\citep{Dalal2002, Koopmans2009, Newman2013, Vegetti2023}.
Additionally, strong lenses offer unique opportunities to study distant, magnified galaxies 
\citep{Lotz2017, Atek2015} 
and constrain the expansion rate of the universe through time-delay Hubble constant measurements of multiply imaged supernovae and quasars 
\citep{Refsdal1964b, Wong2020, Kelly2023, Acebron2023, Pascale2024}.

Existing surveys have already revealed a substantial number of lensing systems, and this count is expected to increase significantly with upcoming imaging surveys. The forthcoming LSST and Euclid surveys are set to usher in a new era in lensing studies; they should discover $120\,000$ and $170\,000$ strong galaxy-galaxy lenses, respectively \citep{Collett2015}. Upon the assumption of the lens distribution as delineated in \cite{Oguri2006}, this projection would correspond to approximately $20\,500$ ($29\,000$) group-scale lenses and $5\,900$ ($8\,300$) cluster-scale lenses within the scope of LSST (Euclid). Coupled with machine learning detection methods \citep{Jacobs2019, Magro2021, Huang2021, Rojas2022, Shu2022, Zaborowski2023}, these surveys will provide an extensive sample of galaxy-to-cluster scale lenses, facilitating statistical applications.

For these applications, accurate, fast, and automated modeling of strong gravitational lenses is essential. The lens model parameterizes the lens mass distribution and source properties, and constraining these parameters to observations is a crucial step. Traditional modeling techniques can be time-consuming, taking hours and even days for cluster-scale lenses to explore the parameter space, making the large number of lenses in forthcoming surveys challenging to handle efficiently. 

This article addresses the specific challenges of achieving fast modeling for group- and cluster-scale lenses. These lenses are commonly modeled using the positions of strongly lensed images, usually knots of star formation in extended arcs \citep{Cerny2018, Sharon2020, Mahler2023, Bergamini2023}. Identifying and pairing multiple image sets requires expertise, which is a challenge for automated modeling. Furthermore, the identification of point sources requires high-resolution imaging, thus limiting the datasets to those from space-based telescopes. Despite these drawbacks, using the positions of these lensed images is fast, making it a preferred method for modeling complex mass distributions. While the use of extended images instead of positions would be ideal for clusters due to the detailed information they provide, the computational time required becomes prohibitive. This is because the cluster mass model involves numerous parameters, and each source requires shape model parameters. 

Pioneering work in automated modeling, such as that by \cite{Zitrin2012} and \cite{Stapelberg2019}, successfully used the light of cluster members to model cluster-scale lenses without relying on the identification of lensed images. While fast, this approach assumes that ``light traces mass'', requiring a calibration sample and thus making it dependent on the selected sample and less sensitive to the system's peculiarities like variations on the mass-to-light ratio and unrelaxed mass distributions.

On the other hand, galaxy-scale lenses are modeled through simulations, reconstructing the surface mass of the lens and the surface brightness of the source and lens galaxy, which are then compared with observations \citep{Birrer2015, Nightingale2018}. Although this approach can be computationally expensive, significant progress has been made in achieving the modeling speed needed for large surveys, employing machine learning aided by graphics processing units (GPU) \citep{Hezaveh2017, Morningstar2019, Pearson2019, Chianese2020, Pearson2021, Schuldt2021, Biggio2022, Adam2023, Schuldt2023}. Among them, GIGA-Lens  \citep{Gu2022}, a promising new software for fast and automated galaxy-galaxy lens modeling, leverages GPU acceleration to achieve comprehensive Bayesian posterior estimates in a few minutes for each system.   

Although this method has been successful for galaxy-scale systems, scaling this method to group or cluster lenses is challenging. 
Cluster lenses often exhibit complex, unrelaxed, and asymmetrical mass distributions \citep{Merten2011, Limousin2012, Jauzac2016}, significantly influenced by subhalos \citep{Meneghetti2007}, often resulting in limited prior knowledge of the specific system.
This added to the large number of parameters needed to describe both the lens and the sources may lead to the ``curse of dimensionality'': the exploration space grows exponentially with the number of free parameters, which, when combined with the weak prior, would require an increasing number of samples to obtain the posterior distribution.

We propose adapting the established methods from galaxy-scale lens modeling to overcome the challenges in modeling lenses at the group and cluster scales, by modifying the GIGA-Lens code with a novel inference approach to handle the increased model complexity. This enhancement involves a hybrid approach that integrates the approximate positions of multiple images with their surface brightness to efficiently estimate the lens mass and source brightness parameters. 

The design is primarily focused on achieving fast inference for these systems using ground-based data as constraints. This is of special interest for lens systems on the LSST survey, as it will have a high depth but low resolution compared to space-based surveys like Euclid. This resolution constraint might hinder the detection of galaxy-scale lenses, but larger-scale lenses remain unaffected. We illustrate the method by applying it to a group-scale strong lens candidate DES J0248-3955 \citep{Jacobs2019} to demonstrate its capability to provide accurate lens models, even with the limited resolution challenges posed by surveys such as LSST.

The system under study is relevant on its own as a potential candidate for having two source planes. \cite{Collett2012} has shown that it is possible to constrain the density of matter $\Omega_{m}$ and the equation of state of dark energy $w$ cosmological parameters independently of $H_0$ with this kind of lens system. While cluster-scale lenses frequently feature multiple source systems, allowing robust cosmological constraints \citep{Jullo2010, Caminha2022}, their complex mass distribution limits their applicability to a few well-studied lenses. Galaxy-scale strong lenses with multiple sources are rarer, but more suitable for statistical application \citep{sharma2023}. 
As a galaxy group lens, DES J0248-3955 is a mid-step between the two: more massive than galaxies but simpler than clusters, making it an interesting candidate to probe cosmology (with future observations) and test the GIGA-Lens code.

The remainder of the paper is organized as follows. Section 2 presents the details of the data. In Section 3, our novel hybrid method is described. Section 4 introduces the lens model for the system. Section 5 presents the resulting lens model and possible sources of errors. Section 6 discusses the advantages and disadvantages of the algorithm.

Magnitudes are reported in the AB system. Our analysis is based on the assumption of a flat $\Lambda$CDM model with $\Omega_{m,0} = 0.3$ and $\text{H}_0 = 70 \, \text{km s}^{-1} \text{Mpc}^{-1}$.


\section{Data} \label{sec:data}
The system under study is a compact group lens candidate listed in \textit{NeuraLens} \citep{Storfer2022}, a catalog of $3057$ strong lens candidates, including galaxies, groups, and clusters. The network assigned a probability of 1.0 that it was a lens, and a human assessment gave it the maximum rating. This system was first discovered by a systematic search with neural networks in DES imaging data by \cite{Jacobs2019}. \cite{Odonnell2022} confirmed it with a rank of 6/10 and measured an Einstein radius of $\theta_E = 3.95''\pm0.40''$ as the average distance between each image and the brightest lens. A photometric redshift $z_{\rm{phot}} = 0.67 \pm 0.02$ was derived from the DES Y6 photometry for the brightest group galaxy (BGG) (g2 in our labeling) located at $02:48:09.54$, $-39:55:48.4$.

\begin{figure}[!htbp]
     \centering
     \begin{subfigure}[b]{0.49\linewidth}
         \centering
         \includegraphics[width=1\textwidth]{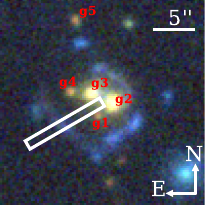}
         \caption{}
         \label{fig:color_img}
     \end{subfigure}
     \hfill
     \begin{subfigure}[b]{0.49\linewidth}
         \centering
         \includegraphics[width=1\textwidth]{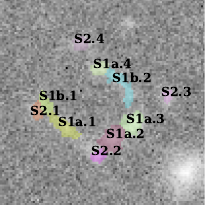}
         \caption{}
         \label{fig:segmentation}
     \end{subfigure}
     \caption{%
     \textbf{(a):} Color composite of the system, the red labels indicate possible group members and the white box represents the X-shooter slit passing through image S1a.1 and between members g2 and g3. 
     \textbf{(b):} g-band image with S\'{e}rsic surface brightness model of g1 to g5 subtracted, overlaid with the segmentation of the multiple images labeled in black.}
\end{figure}

We used the \textit{grz}-band DECam images from the DESI Legacy Imaging Survey data release 9 \citep{Dey2019}, to produce a composite color image shown in Fig.~\ref{fig:color_img}. 
We identified the central group members (g1-g5) based on the $g-r$ and $r-z$ colors shown in Table~\ref{tab:photometry}. 
We modeled their surface brightness with the same code used for modeling the lens, but without incorporating lens deflection. Each galaxy was modeled with a single S\'{e}rsic profile, and the photometric magnitudes derived from the model are reported in Table~\ref{tab:photometry}. 
We also provide the photometric redshifts from \cite{Zhou2021}. These were calculated using a random forest regression algorithm that incorporates photometric data in the \textit{grz} bands, as well as \textit{W1} and \textit{W2} bands, in addition to information on the morphology of the galaxies.

The \textit{g}-band image after g1-g5 surface brightness subtraction is displayed in Fig.~\ref{fig:segmentation}, featuring the segmentation of multiple-image candidates obtained with Photutils software \citep{Bradley2023}. The image labeling has a format $Sij.k$ being $i$ the source plane, $j$ the identifier of a family of images from the same source in that plane, and $k$ the identifier of the image within the family.

Spectroscopic data were acquired from the X-shooter instrument mounted on the VLT-UT3 \citep{Vernet2011}, through the observation program ID 110.23U2.001 (PI: Jullo). This program aims to characterize the NeuraLens selection function by measuring the redshift of 100 lens systems (Jullo et al. in prep). The observation was made with an air mass of 1.041 and a seeing of about 1.75''. The time spent on target was divided into exposures of 2x1468s, 2x1384s, and 5x300s each in the UBV, VIS, and NIR channels respectively. Each exposure was reduced independently with the ESO/REFLEX pipeline in STARE mode. Sky lines were template subtracted and the sky residual present in a fraction of the slit was subtracted to produce the spectra presented in Fig.~\ref{fig:spectrum}. 

As depicted in Figure~\ref{fig:color_img}, the slit passes through the east arc S1a and close to the group members g2 and g3. As shown in Figure~\ref{fig:spectrum}, the source spectrum presents a clear $\rm{H}_\alpha$ emission line at redshift $z_{S1a}=1.2722 \pm 0.0005$. An [O II] doublet is also detected at this redshift but is highly affected by skylines. The 1D spectrum of the source was obtained by optimal spectral extraction \citep{Horne1986}, with a spatial profile obtained with an aperture of $12 \, \text{\AA}$ around the $\rm{H}_\alpha$ emission.

The one group member with a spectroscopic redshift has a spectrum that does not show any emission lines. A redshift of 
$z_{BGG,spec} = 0.685 \pm 0.002$
was deduced from the K and H absorption lines. This spectrum is associated with members g2 and g3, likely dominated by g2 as the BGG. The offset between the dark matter halo and the brightest galaxy is expected to be in the order of a few kpc according to simulations \citep{Roche2024} and observations \citep{Harvey2017} meaning that the BGG is representative of the group redshift. This is supported by the photometric redshifts of the group members, as shown in Table~\ref{tab:photometry}, with the variance-weighted average redshift of the central galaxies g1-g4 being $z_{L-phot} = 0.67 \pm 0.04$. The $4\,000 \, \text{\AA}$ break falls within the r band, allowing the g, r, and z bands to bracket the break and determine their redshift.

For the redshift of the group, we adopt the spectroscopic redshift of the BGG as its centroid. Given that this redshift is based on a single galaxy, we account for the uncertainty in the group redshift by incorporating the standard error of the variance-weighted average of the photometric redshifts of the central members. This approach yields an estimated group redshift of $z_{L} = 0.69 \pm 0.04$. While this approach incorporates potential variations within the group, additional observations are required to confirm the accuracy of the group redshift, given the limited number of spectroscopic measurements currently available.

\begin{table}[!htbp]
    \caption{The table shows photometric redshifts from \cite{Zhou2021}. The spectroscopic redshift of the BCG (g2), $z_{BGG,spec} = 0.685 \pm 0.002$ is consistent with the photometric one $z_{BGG,phot} = 0.68 \pm 0.06$. The colors, obtained with a $1''$ diameter circular aperture, and the z-band magnitude, computed through S\'ersic models, are also exhibited.}
    \label{tab:photometry}
    \centering
    \begin{tabular}{c|c|c|c|c}
         Member & photo z & $g - r$ & $r - z$ & $z$ \\
         \hline
         g1 & 0.6 $\pm$ 0.2 & 1.63 & 1.69 & 21.80 \\
         g2 & 0.68 $\pm$ 0.06 & 1.80 & 1.74 & 18.79 \\
         g3 & 0.66 $\pm$ 0.06 & 1.58 & 1.64 & 19.11 \\
         g4 & 0.8 $\pm$ 0.3 & 1.88 & 1.69 & 21.18 \\
         g5 & 0.66 $\pm$ 0.07 & 1.74 & 1.76 & 21.04 \\
    \end{tabular}
\end{table}

\section{Method} \label{sec:method}
Gravitational lensing arises from the bending of the light path in curved spacetime, a phenomenon described by the lensing equation \citep{Schneider1992}:

\begin{equation} \label{eq:lens_equation} 
    \beta(\theta) = \theta - \alpha
\end{equation}

\noindent where $\beta$ is the angular position in the source plane, $\theta$ in the image plane, and $\alpha$ is the deflection angle due to the lens at the image position. In the context of strong lensing, this equation yields multiple solutions that account for multiple images. The deflection is related to the convergence $\kappa(\theta)$ through:

\begin{equation} \label{eq:deflection} 
    \alpha(\theta) = \frac{1}{\pi} \int \kappa(\theta') \frac{\theta - \theta'}{|\theta - \theta'|^2} d \theta'^2
\end{equation}

\noindent where $\kappa(\theta)$ is the projected mass density of the lens $\Sigma(\theta)$ over the critical density $\Sigma_{crit}$, the latter dependent on the angular diameter distances between the observer and the lens ($D_L$), the observer and source ($D_S$) and the lens and source ($D_{LS}$):

\begin{equation}\label{eq:critical_density} 
    \Sigma_{crit} \equiv \frac{c^2 D_S}{4\pi G  D_L D_{LS}}
\end{equation} 

This means that two sources at different distances are deflected differently by the same lens. To be able to model sources at different redshifts $z_1$ and $z_2$, we convert $\alpha$ from one source plane to another by rescaling it by the distance ratio $\eta$, defined as:

\begin{equation} \label{eq:distance_ratio} 
    \frac{\alpha_1}{\alpha_2} = \frac{D_{LS}(z_{1}) D_{S}(z_2)}{D_{S}(z_1) D_{LS}(z_{2})} \equiv \eta
\end{equation}

We use a parametric modeling approach, that is, the lens potential is described by an analytic profile dependent on a set of parameters $\Theta$. These parameters can be inferred by constraining them to observations. In cluster lens modeling, the main constraint is the position of the multiple images. The model parameters can be inferred by minimizing the $\chi^2$ between the observed image positions and the ones predicted by the model \citep{Schneider1988}:

\begin{equation} \label{eq:chi2_ip_positions}  
    \chi^2_{\theta} = \sum_{i=1}^{N} \sum_{j=1}^{n_i} \frac{\left( \theta_{obs, j} - \theta_{model, j}(\Theta) \right)^2}{\sigma^2_{\theta, j}}
\end{equation}

\noindent for a system comprising $N$ sources, each source $i$ exhibits $n_i$ images with positions $\theta_j$ and associated errors $\sigma^2_{\theta, j}$.

This approach faces two primary challenges: matching observed and predicted images, and the computationally expensive task of inverting the lens equation. A viable alternative is to employ a first-order approximation \citep{Kochanek1991}, $\delta \beta \approx \mu^{-1} \, \delta \theta$, which allows us to compute $\chi^2$ in the source plane:

\begin{equation} \label{eq:chi2_sp_positions}  
    \chi^2_{\beta} = \sum_{i=1}^{N} \sum_{j=1}^{n_i} \frac{\left( \beta(\theta_{obs, j}; \Theta) - \beta_{model}(\Theta) \right)^2}{\mu^{-2}(\theta_{obs,j}; \Theta) \; \sigma^2_{\theta, j}}
\end{equation}

where $\beta(\theta_{obs, j}; \Theta)$ is the position of the observed image $j$ mapped to the source plane and $\mu$ is its magnification. This is straightforward to compute, as it avoids the need to invert equation~(\ref{eq:lens_equation}). While most elliptical mass profiles lack an analytical expression for $\mu$, our code efficiently calculates it using automatic differentiation, ensuring accurate gradients and simplifying the computation.

As the $\chi^2$ given by equations (\ref{eq:chi2_ip_positions}) and (\ref{eq:chi2_sp_positions}) depend on the image positions, they only give a good constraint to the model when point-like features can be identified in the images, thus, this is only possible with high-resolution data.

In the case of galaxy-scale lens modeling, $\chi^2$ is computed with the pixel-to-pixel comparison between the surface brightness of the observed image $\mathcal{I}_{obs}$ (in units of counts/s) and a simulation $\mathcal{I}_{model}$. For this, the model needs to include not only the source position but also its surface brightness, and the $\chi^2$ is defined as \citep{Birrer2015}:

\begin{equation} \label{eq:chi2_pixels}  
    \chi^2_{pix} = \sum_{x,y} \frac{\left( \mathcal{I}_{obs}(x, y) - \mathcal{I}_{model}(x, y; \Theta) \right)^2}{\sigma^2_{pix}(x, y; \Theta)}
\end{equation}

The error $\sigma^2_{\text{pix}}$ includes contributions from background noise and Poisson noise, related to exposure time $t_{\text{exp}}$ and gain $\mathcal{G}$:

\begin{equation} \label{eq:sigma}
    \sigma^2_{pix} = \sigma^2_{bkg} + \sum_{x,y}\frac{\mathcal{I}_{model}(x, y; \Theta)}{\mathcal{G} \, t_{exp}}
\end{equation}

This method offers robust constraints on extended images such as arcs and Einstein rings in galaxy-galaxy lensing, and has proved to be reliable on ground-based data \citep{Knabel2023, Schuldt2023}.

Parameter estimation is accomplished through a Bayesian method, assuming a normal distribution for the likelihood:

\begin{equation} \label{eq:likelihood}
    log \, \mathcal{L} = -\frac{1}{2} \left[ \chi^2 + log \left( 2 \pi 
 \sigma^2 \right)\right]
\end{equation}

This requires the choice of a prior on the parameters $\pi(\Theta)$, and exploring the posterior distribution $\mathcal{P}(\Theta; \mathcal{D}) = \pi(\Theta) \, \mathcal{L}$ possibly in multiple steps.

The Giga-Lens code \citep{Gu2022} relies on equation (\ref{eq:chi2_pixels}) to obtain the posterior in three steps. 
First, the parameters that yield the maximum probability (maximum a posteriori, MAP) are obtained by a multi-start gradient descent. This method involves executing multiple optimization trials from different initial parameter values to reduce the chance of getting stuck in local minima.
Then, the posterior distribution is approximated with a Normal distribution centered on the MAP estimate using variational inference. This method approximates complex probability distributions by optimizing a more straightforward distribution, thereby improving computational efficiency.
Finally, the posterior distribution is obtained by the Markov Chain Monte Carlo (MCMC) method using the Hamiltonian Monte Carlo (HMC) kernel, initialized by the distribution obtained in the previous step. The three steps use gradient information obtained by automatic differentiation to guide the exploration efficiently. 
Although this method has proven to be fast in galaxy-galaxy scenarios, the multi-start gradient descent stage may fail to obtain the MAP in the case of a large exploration space, leading to a biased result.

In this paper, we propose an alternative exploration method to account for the complexity of multiple-galaxy lens systems. We primarily rely on equation (\ref{eq:chi2_pixels}) as the main constraint for our model. However, instead of simulating the entire pixel matrix, we limit it to a region of interest, typically defined by a mask that encompasses the multiple images (as depicted in Fig.~\ref{fig:segmentation}). During the model evaluation, the pixel grid is represented by a sparse matrix instead of a dense one. This approach is more memory-efficient and avoids the need to simulate the surface brightness of cluster members or field galaxies distant from the images. 

The minimization of the equation (\ref{eq:chi2_pixels}) can be slow initially, as the simulated images can deviate significantly from the observed ones, resulting in a flat $\chi^2$ surface. To address this, we initially employ equation (\ref{eq:chi2_sp_positions}) to approximate the high-probability region and subsequently refine the solution using equation (\ref{eq:chi2_pixels}). This exploration step is much faster as the $\chi^2$ decreases when the images are mapped closer in the source plane, leading to a convex surface in the same situation when the other is flat. Additionally, it does not require computing the surface brightness of the deflected source, this leads to a much faster likelihood evaluation.

To join both steps we employ a simulated annealing approach by weighting each likelihood term by a power $\lambda$ and gradually transitioning from the prior $\pi(\Theta)$ to the posterior $\mathcal{P}(\Theta; \mathcal{D})$ by varying $\lambda$ in two stages:

\begin{equation} \label{eq:posterior}
    \mathcal{P}_{\lambda_1,\lambda_2}(\Theta; \mathcal{D}) = \pi(\Theta) \, \mathcal{L}_{\beta}^{\lambda_1} \, \mathcal{L}_{pix}^{\lambda_2} 
\end{equation}

\noindent where $\mathcal{L}_{\beta}$ is the likelihood given by the source plane positions (equation (\ref{eq:chi2_sp_positions})), and $\mathcal{L}_{pix}$ by pixels (equation (\ref{eq:chi2_pixels})). Initially, we set $\lambda_2 = 0$ and increase $\lambda_1$ from 0 to 1, this way, we start by sampling the prior and increase the weight of $\mathcal{L}_{\beta}$ to approach the posterior constrained by the position of the images. Subsequently, we increase $\lambda_2$ to 1 and set $\lambda_1 = 1 - \lambda_2$ to slowly mutate the previous posterior to the one constrained by pixels. Finally, we keep sampling with fixed $\lambda_1 = 0$ and $\lambda_2 = 1$ to obtain accurate probability contours. The prior remains unchanged throughout the sampling process.

The sampling is performed using a Sequential Monte Carlo (SMC) method available through TensorFlow Probability \citep{TensorFlowProb2017}, heavily based on \cite{DelMoral2012}. The posterior is modeled with a set of weighted ``particles'' which are updated using an MCMC transition kernel and resampled to generate a new set of particles. For the MCMC step, we employ the HMC kernel, which uses gradient-informed evolution through an analog of Hamiltonian dynamics, enabling the exploration of complex posteriors with minimal likelihood evaluations. In the first phase, the particles are initialized from a random sample from the prior of the lens mass. In the second one, we take a subsample of the particles resulting from the previous step and extend their dimension with a random sample from the source light prior. 

The code is implemented in JAX \citep{jax2018github}, a library for high-performance machine learning that enables GPU parallelization and efficient computation of gradients through automatic differentiation. Additionally, we employ TensorFlow Probability \citep{TensorFlowProb2017}, a JAX-compatible package for probabilistic programming. This allows us to compute multiple simulated images simultaneously and handle the complex optimization and sampling processes effectively.

\section{Lens Model} \label{sec:model}
The adopted lens model corresponds to a singular isothermal ellipsoid (SIE) mass profile that describes the mass of the group. The radial SIE convergence is given by:

\begin{equation} \label{eq:sie}
    \kappa(r) = \frac{1}{2} \frac{\theta_E}{r} 
\end{equation}

\noindent where $\theta_E$ is the Einstein radius of the lens relative to a source at a distance $D_S$, which relates to the velocity dispersion of the deflector as:

\begin{equation} \label{eq:sigma_v}
    \sigma_v = \sqrt{\frac{\theta_E c^2}{4 \pi}\frac{D_S}{D_{LS}}}
\end{equation}

We also include a shear component that describes the effect of external structures.

Each source galaxy surface brightness is described with a S\'ersic profile. Consequently, the model encompasses a total of 7 free parameters for the lens mass and 7 parameters per source galaxy.

The lens model is constrained using the foreground-light subtracted g-band image, limited to the pixels within the segmentation image depicted in Fig.~\ref{fig:segmentation}, with an additional extension of 3 pixels. This ensures that we include adjacent pixels and prevents the model images from extending further than the observed ones. Regarding the image position, we use the flux-weighted centroid of the brightest 20\% pixels within the segment and its uncertainty. 

Due to large errors in the colors, which hindered an accurate match of the multiple images, we employed the model to predict the image families. For this, we explored various models considering a single source and different image combinations until we obtained a model that successfully reproduces images S1a.1-4, from a single source, denoted S1a, located at the spectroscopic redshift $z_{S1} = 1.2722$. 

We then employ this model to predict the counterimages of the remaining 1 to 2 image systems by mapping each image to the source plane and then back to the image plane assuming an arbitrary redshift. With this method, we predict the presence of a second source, denoted S1b, also at $z_{S1}$, which contributes to the images S1b.1-2, and a third source, labeled S2, at a higher redshift ($z_{S2} \approx 3$), generating a quadruple image pattern comprising images S2.1-4.

The conclusive model incorporates all three aforementioned sources and introduces an additional parameter $\eta$ that accounts for the distance to source S2 (see equation (\ref{eq:distance_ratio})). This yields a total of 29 independent parameters, which collectively define the model's complexity and flexibility. We explore the parameter space with the pipeline described in Sect.~\ref{sec:method} using $1\,000$ particles for the SMC, and then making 100 sampling steps making a total of $100\,000$ samples.

We assume non-informative uniform priors for all the parameters describing the lens mass distribution, the positions of the sources and $\eta$; and normal priors for the parameters describing the S\'ersic profiles. The complete prior is shown in Table~\ref{tab:prior}. This prior is highly flexible, covering a broad spectrum of lenses and sources. It is important to note that the same prior applies to all three sources, except for the ellipticity of S2, which is more tightly constrained due to the minimal elongation observed in the four images.

\begin{table}[ht]
    \centering
    \caption{Prior distribution for the lens parameters.}
    \label{tab:prior}
    \begin{tabular}{c}
    \hline 
    \hline \\
    \begin{equation*}
    \begin{aligned}
    \text{Lens mass:} & 
    \begin{cases}
    \theta_E &\sim \mathcal{U}(3, 8) \\
    x, y &\sim \mathcal{U}(-2, 2) \\
    e_1, e_2 &\sim \mathcal{TN}(0.0, 0.2; -0.5, 0.5) \\
    \end{cases} \\
    \text{Source light:} & 
    \begin{cases}
    R &\sim \mathcal{N}(0.4, 0.1) \\
    n &\sim \mathcal{TN}(3.0, 2.0; 0.5, 5.0) \\
    I &\sim \mathcal{N}(0.25, 0.07) \\
    x, y &\sim \mathcal{U}(-3, 3) \\
    e_1, e_2 &\sim \mathcal{TN}(0.0, 0.2 \, (0.05); -0.5, 0.5) \\
    \end{cases} \\
    \text{Source distance:} & 
    \begin{cases}
    \eta \qquad \! &\sim \mathcal{U}(0.6, 0.7) \\
    \end{cases} \\
    \end{aligned}
    \end{equation*}\\
    \hline
    \end{tabular}
    \tablefoot{
    The lens mass profile is parametrized by the Einstein radius $\theta_E$, and the source surface brightness by the S\'{e}rsic radius $R$, index $n$, and half-light $I$. For both cases, the elliptical geometry is described by its center $(x, y)$ and eccentricities $e_1$, $e_2$. $\mathcal{U}(a, b)$ denotes a uniform distribution, $\mathcal{N}(\mu, \sigma)$ a normal distribution, and $\mathcal{TN}(\mu, \sigma; a, b)$ a truncated normal distribution. The value in parentheses is a different prior for S2 eccentricities as is expected to have low elongation. The source distance prior limits $z_{S2}$ to $(2.1, 3.2)$. All positions and radius ($x$, $y$, $\theta_E$ and $R$) are expressed in arcseconds, and the positions are relative to the central point of the pixelated image. $I$ is in micro Jansky per suqared arcsecond. A normal prior on $e_1$ and $e_2$ is needed to reflect a rotational symmetry (see \cite{Gu2022}).}
\end{table}

The final model is shown in Fig.~\ref{fig:model_final} and is discussed in further detail in the following section. 

\section{Results} \label{sec:results}

\subsection{Model Constraints and Possible Sources of Errors}

\begin{figure*}[!t]
    \centering
    \includegraphics[width=\textwidth]{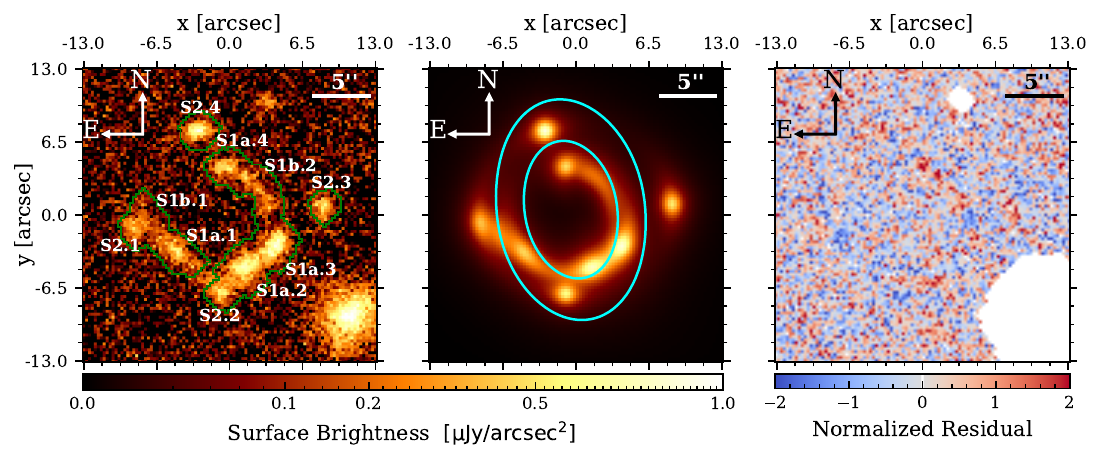}
    \caption{%
    \textbf{Left panel:} g-band image with g1-g5 foreground-light subtracted. The green regions show the pixels included in the simulation. Image systems are labeled as S1a.1-4, S1b.1-2, and S2.1-4.
    \textbf{Middle panel:} Model reproduction of the image based on the median of the marginalized parameters. The tangential critical lines are shown in cyan for the source plane S1 (inner) and S2 (outer).
    \textbf{Right panel:} Normalized residual between the model and the g-band image. The remaining foreground galaxies are masked; the residual is below $2 \sigma$ for most of the model.     
    }
    \label{fig:model_final}
\end{figure*}

\begingroup
\renewcommand{\arraystretch}{1.5}
\begin{table*}[!t]
    \centering
    \caption{Median values and $1\sigma$ confidence intervals for the model parameters of the lens model. The parameters include the Einstein radius ($\theta_E$), right ascension, declination, position angle ($\phi$), and ellipticity ($\epsilon$); the external shear components ($\gamma_1$ and $\gamma_2$); and the distance ratio between sources S1 and S2 ($\eta$) as defined in Equation~(\ref{eq:distance_ratio}). The parameters $\theta_E$, $\gamma_1$, and $\gamma_2$ are relative to source S1 at redshift $z=1.2722$.}
    \label{tab:model}
    \begin{tabular}{ccccc|cc|c}
         \hline
         \hline 
         \multicolumn{5}{c|}{Halo} &
         \multicolumn{2}{c|}{Shear} &
         \multicolumn{1}{c}{S2} \\
         \hline 
         $\theta_{E}$ & $RA$ & $DEC$ & $\phi$ & $\epsilon$ & $\gamma_1$ & $\gamma_2$ & $\eta$ \\
         $[arcsec]$ &  &  & $[deg]$ &  &  &  &  \\
         $5.052^{+0.005}_{-0.005}$ & $2h48m10s$ & $-39d55m48s$ & 
         $12.1^{+0.3}_{-0.2}$ & $0.203^{+0.002}_{-0.002}$ & 
         $0.003^{+0.001}_{-0.001}$ & $-0.015^{+0.001}_{-0.001}$ & $0.626^{+0.001}_{-0.001}$  \\
         \hline 
    \end{tabular}
\end{table*}
\endgroup

Despite the simplicity of the mass profile, the model shown in Fig.~\ref{fig:model_final} produces a suitable reproduction of the observed configuration of image systems S1a, S1b, and S2 within the errors. The reduced $\chi^2$ is $\chi^2_{\nu,pix} = 1.0$ within the simulated region (enclosed in green in Fig.~\ref{fig:model_final}), and $\chi^2_{\nu,pix} = 1.1$ when only considering pixels with a signal-to-noise greater than 1 (the same as in Fig.~\ref{fig:segmentation}). 
The normalized residual shown in Fig.~\ref{fig:model_final} indicates that the reproduction of the S1a images is at the noise level. In contrast, the images of S1b are not as accurately reproduced as those of S1a. Specifically, S1b.2 displays three bright clumps that exceed the model predictions. This discrepancy could be due to the magnification effects of adjacent group members g1 and g2, or to a source complexity that a smooth S\'ersic profile does not capture. However, the model replicates the shape of the arc.
The four images of S2 are well reproduced by the model, favoring the hypothesis that it is a strongly lensed source at higher redshift. The residual slightly exceeds that of S1a, attributed to a positional offset of the images, with a root mean square error of $0.22''$ in the image plane. The offset may be due to the influence of the source being doubly lensed, first by S1 and then by the group (see below). 

As indicated in Table~\ref{tab:model}, the model appears highly constrained and the uncertainties in the lens parameters are likely underestimated. For example, the Einstein radius $\theta_E$ has an uncertainty of only $0.1\%$, which is extremely low considering the low signal-to-noise ratio and resolution of the data. 
Although the simple SIE model for the halo and S\'ersic profiles for the sources effectively mimic the data, their rigidity may cause only a limited portion of the parameter space to replicate the observations. This problem might be mitigated by adopting a more sophisticated model for the lens mass, such as an elliptical power law with a free slope, or by including a mass model for the group members.

The model has a small but non-negligible shear of $\gamma_{ext} = 0.015$, which can be attributed to the faint group members that extend further than the center described by the model, as well as a richer galaxy group located approximately 2' to the south.

\subsection{Lens and Sources Analysis}

\begin{figure}[!htb]
    \centering
    \includegraphics[width=1\linewidth]{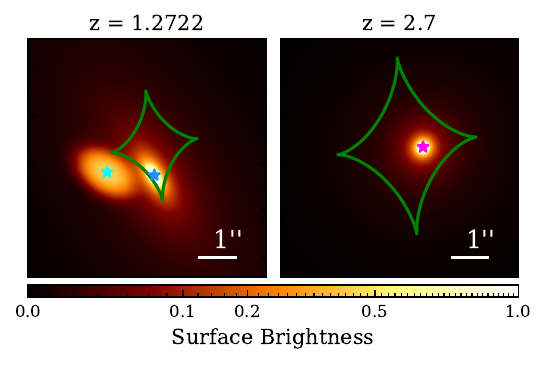}
    \caption{Surface brightness model for the sources S1a and S1b (left) and S2 (right) in micro Jansky per suqared arcsecond. The green lines show the caustic. The blue, cyan, and magenta stars indicate the point source at the center of S1a, S1b, and S2 respectively. Source S2 is predicted to be at $z = 2.7$ by the single-lens plane model.}
    \label{fig:sources}
\end{figure}

As can be seen in Table~\ref{tab:model}, our measurement of the Einstein radius for the source S1 is considerably larger than that of \cite{Odonnell2022} $\theta_E = 3.95''\pm0.40''$ measured with the positions of the images. The difference is probably due to the high ellipticity of the halo in our model $\epsilon = 0.203 \pm 0.002$ (defined as $\epsilon = (1 - q)/(1 + q)$ with $q$ the axis ratio), which deviates largely from a spherical case ($\epsilon = 0$) assumed by \cite{Odonnell2022}. Using this Einstein radius and the equation (\ref{eq:sigma_v}) we measure a velocity dispersion for the galaxy group of $\sigma_v = (690 \pm 30) \, km \, s^{-1}$, considering the uncertainties associated with the model as well as the redshifts of the source and lens.

The distance ratio between the source planes S1 and S2 deviates from unity with a discrepancy greater than $5 \sigma$. This deviation persists even when accounting for a more conservative uncertainty of 10\%, yielding $\eta = 0.63 \pm 0.06$. This provides strong evidence for the system having two distinct source planes, thereby capable of constraining cosmology. 
According to \cite{Collett2012}, the ratio of the deflection angles $\eta$ is a measure of the distance ratios (see equation~(\ref{eq:distance_ratio})), which can be compared with a redshift measure to determine the density of matter $\Omega_{m}$ and the equation of state of dark energy $w$. To do this, spectroscopic observations of S2 are necessary, which may be the focus of a future study.

Furthermore, this would require a double-lens-plane model; otherwise, the single-plane model would overestimate the distance to S2 by not accounting for the additional deflection induced by S1 (see \cite{Collett2014}). 
In the single-plane model, the light is deflected only once, projecting the source further away than in the double-plane case, where the second deflection allows the source to be closer while still matching the observed image separation. 
Thus, with our cosmological assumptions and disregarding the mass of S1a, the model estimates an upper limit for the redshift of S2 to be $z_{S2} \leq 2.7$. Given that the mass of S1 is significantly lower than that of the group, the redshift of S2 is anticipated to approach this upper boundary.

The source surface brightness models for both source planes are shown in Fig.~\ref{fig:sources}. There are two sources in the left panel: S1a, which has a spectroscopic redshift of $z = 1.2722$, and S1b, which the model predicts to be at a similar redshift, and we fix its redshift to $z = 1.2722$ during modeling. The difference in color between S1a ($g - r = 0.19 \pm 0.01$) and S1b ($g - r = 0.31 \pm 0.24$) suggests they are different galaxies, although spectra of S1b and higher-quality imaging would be needed to confirm this. The right panel shows the third source, S2, which the model predicts to be at approximately $z \sim 2.7$.

\subsection{Impact of S1a Mass on S2 Images}

Our algorithm currently does not support multiple lens planes, but we have explored the impact of S1 on S2 using Lenstronomy \citep{Birrer2018, Birrer2021}. We performed a double lens plane simulation with an added SIE mass profile for S1a, mirroring its light profile shape. We calculated S1a's velocity dispersion using a mass-to-light ratio (M/L), assuming the lowest reported ratio ($M/L \sim 0.3$) by \cite{VandeSande2015}. This yields a dynamical mass of about $6.4 \times 10^{9} \, M_\odot$ for S1a, which leads to a velocity dispersion of $\sigma_v \sim 100 \, km\,s^{-1}$ when asumming an isothermal profile.

\begin{figure}[!htb]
    \centering
    \includegraphics[width=\linewidth]{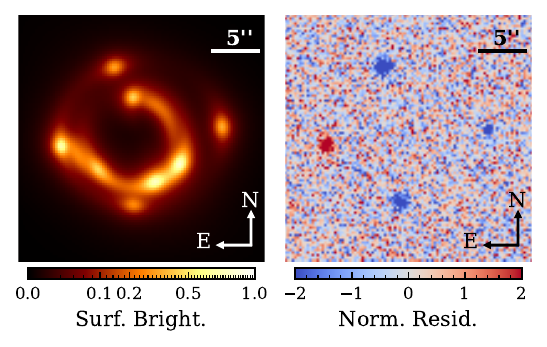}
    \caption{\textbf{Left:} The model consists of two lens planes. The plane at $z_{L}$ has the same mass model as the one shown in Fig.~\ref{fig:model_final}, while the plane at $z_{S1}$ is characterized by a SIE model with a velocity dispersion of $\sigma_{v, S1} = 100 \, km/s$ and the same geometry as the surface brightness profile of S1a. \textbf{Right:} The normalized residual between the dual and single plane models, along with the noise, is shown. The S2 images exhibit a noticeable difference between the two models, indicating that the mass of S1a has a significant impact.}
    \label{fig:multiplane}
\end{figure}

As shown in Fig.~\ref{fig:multiplane}, the mass of S1a changes the magnification of the S2 images and produces a small offset (about $0.12''$). This suggests a substantial lensing impact of S1a on S2 when the dispersion is at least $100 \, km\,s^{-1}$, detectable with a $5\sigma$ significance. Thus, accurate modeling of S2 requires a multiplane lens approach, which we aim to incorporate in our forthcoming research.

\section{Discussion} \label{sec:discussion}

Compared to the original GIGA-Lens modeling \citep{Gu2022}, our algorithm has several advantages when dealing with complex lens models or large pixelated images. The memory requirements are greatly reduced when modeling only the region of interest. In our case, this reduces the number of simulated pixels from $10\,000$ to $1\,506$ during profile evaluation. With this number of pixels, we can simulate $\sim 10\,000$ models and their gradients simultaneously on an NVIDIA A100 GPU. Furthermore, while for this system we previously subtracted the foreground light, this step may be omitted when the images are far from the deflector light, simplifying the modeling pipeline. 

We decided to use a multistep annealing algorithm instead of the standard GIGA-Lens inference pipeline. The motivation for this is the following: the GIGA-Lens convergence is highly dependent on the first step, a multi-start gradient descent. If the MAP is not found in this step, the following exploration steps will be biased towards a local maximum. Commonly, this is not a problem for galaxy-scale lenses: the model can be described by a smaller set of parameters and tighter priors. \cite{Gu2022} showed that 300 samples are sufficient for consistent MAP identification. For more complex models, this is not the case, as the number of needed samples increases exponentially with the number of parameters. For example, in our system, we were unable to obtain a consistent result with this algorithm when using $10\,000$ samples (the memory limit with our GPU).

\begin{figure}[!htbp]
     \centering
     \includegraphics[width=1\linewidth]{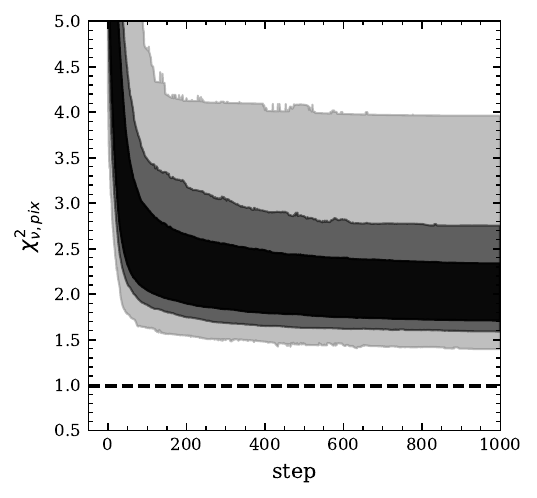}
     \caption{%
     Evolution of the reduced $\chi^2_{pix}$ along multi-start gradient descent steps. The regions account for the $1\sigma$, $2\sigma$, and $3\sigma$ percentiles. The dashed line shows the $\chi^2_{\nu,pix}$ achieved with SMC for comparison.}
     \label{fig:MAP_steps}
\end{figure}

Instead, with the annealing algorithm in SMC, the whole prior is explored if $\lambda$ in the equation (\ref{eq:posterior}) is changed slowly enough as the exploring space is reduced in a manner similar to nested sampling \citep{Skilling2004}. We found that we can achieve consistent results in different runs when using SMC with $1 \, 000$ particles. Furthermore, as demonstrated in Figure~\ref{fig:MAP_steps}, most of the gradient descent samples become stuck after 100 steps and the lower $\chi^2_{\nu,pix}$ is reached at 250 steps, which implies that they quickly settle into the closest local $\chi^2_{\nu,pix}$ minimum, which has a higher $\chi^2_{\nu,pix}$ that the one from SMC. 

An additional advantage of our method is the use of positional information to effectively shrink the parameter space. By leveraging image positions, we significantly reduce the prior volume. For example, this approach allows us to constrain the Einstein radius to $\theta_E = 4.9'' \pm 0.2''$, effectively decreasing the prior variance of $2.1''$ (uniform between $3''$ and $8''$) by 98\% in a few seconds. Consequently, pixelated modeling can then be focused around this refined value. This targeted approach improves the efficiency of our algorithm, particularly in complex scenarios where the parameter space can be large.

Our exploration method not only offers depth but is also fast and efficient. The sampling takes 4.5 minutes, which is comparable to the GIGA-Lens time for a galaxy-scale lens of about 6 minutes using a much stricter prior and half the number of parameters \citep{Gu2022, Cikota2023}. 
As shown by \cite{Gu2022}, this time may be reduced by a factor of $\sim 3.5$ when using 4 A100 GPUs instead of only one.

In this paper, we presented the modeling of a group-scale lens, but larger-scale systems can be modeled using scaling relations for group/cluster members so that the number of parameters does not depend on the number of deflectors, which is commonly assumed for cluster lenses \citep{Brainerd1996, Limousin2005, Jullo2007}. We intend to incorporate this feature in the near future.

\section{Conclusions} \label{sec:conclusions}

We present a novel lens modeling software, which is an improvement on the original GIGA-Lens for simulating complex systems. This enhanced version employs an improved two-step inference method, using image positions and efficient SMC sampling for comprehensive parameter space exploration, even with limited prior knowledge.

We demonstrate our method by modeling the DES J0248-3955 lens system, and the primary findings related to this system are as follows:

\begin{enumerate}
    \item Based on VLT x-shooter spectra, we determine the redshift for the group to be $z_{L} = 0.69 \pm 0.04$, with this value derived from the spectroscopic redshift of the BGG and the photometric redshift estimates of four additional group members. For one of the sources, we measure a redshift of $z_{S1} = 1.2722 \pm 0.0005$.

    \item The observed multiple-image configuration was successfully reproduced using a singular isothermal ellipsoid profile and external shear.
    
    \item With our modeling technique, we obtained the posterior for the model with 29 parameters within a few minutes, with even faster processing possible for simpler models or using multiple GPUs.
    \item  This system is consistent with being a double source plane lens, however, to fully describe the second source, it is necessary to take into account the mass effect of the first source.
    \item Using the lens model, we measured velocity dispersion $\sigma_v = (690 \pm 30) \, km \, s^{-1}$ for the galaxy group.
\end{enumerate}

As a future scope, this lens system holds the potential to constrain the nature of dark energy, which is pending further observational data as well as a double lens plane model.

In conclusion, our software addresses the challenge of modeling strong lenses at group scales using ground-based data by enabling fast modeling and providing strong model constraints even with low-resolution imaging and a broad prior on the lens parameters.

Our software not only improves the modeling of individual systems like DES J0248-3955 but also holds significant potential for broader applications in future large-scale surveys like LSST, which will discover numerous such systems, though follow-up observations for all of them will not be feasible. The use of a broad prior enables the modeling of a diverse range of systems, making it suitable for automated applications. A future study will focus on exploring the scalability and automation potential of our software, further establishing its utility for high-throughput lens modeling.

\begin{acknowledgements}
    Based on data obtained from the ESO Science Archive Facility with DOI(s):
    \url{https://doi.eso.org/10.18727/archive/71}.
    
    The authors acknowledge funding from ANID through BECAS/MAGISTER NACIONAL 22232142, ANID BASAL project FB210003, and FONDECYT project 23050313. We also acknowledge funding from the Centre National de Recherche Scientifique (CNRS) and the Franco-Chilean Laboratory of Astronomy (FCLA), IRL 3386.
    
    The DESI Legacy Imaging Surveys consist of three individual and complementary projects: the Dark Energy Camera Legacy Survey (DECaLS), the Beijing-Arizona Sky Survey (BASS), and the Mayall z-band Legacy Survey (MzLS). DECaLS, BASS and MzLS together include data obtained, respectively, at the Blanco telescope, Cerro Tololo Inter-American Observatory, NSF’s NOIRLab; the Bok telescope, Steward Observatory, University of Arizona; and the Mayall telescope, Kitt Peak National Observatory, NOIRLab. NOIRLab is operated by the Association of Universities for Research in Astronomy (AURA) under a cooperative agreement with the National Science Foundation. Pipeline processing and analyses of the data were supported by NOIRLab and the Lawrence Berkeley National Laboratory (LBNL). Legacy Surveys also uses data products from the Near-Earth Object Wide-field Infrared Survey Explorer (NEOWISE), a project of the Jet Propulsion Laboratory/California Institute of Technology, funded by the National Aeronautics and Space Administration. Legacy Surveys was supported by: the Director, Office of Science, Office of High Energy Physics of the U.S. Department of Energy; the National Energy Research Scientific Computing Center, a DOE Office of Science User Facility; the U.S. National Science Foundation, Division of Astronomical Sciences; the National Astronomical Observatories of China, the Chinese Academy of Sciences and the Chinese National Natural Science Foundation. LBNL is managed by the Regents of the University of California under contract to the U.S. Department of Energy. The complete acknowledgments can be found at \url{https://www.legacysurvey.org/acknowledgment/}.
    
    The Photometric Redshifts for the Legacy Surveys (PRLS) catalog used in this paper was produced thanks to funding from the U.S. Department of Energy Office of Science, Office of High Energy Physics via grant DE-SC0007914.
\end{acknowledgements}

\bibliographystyle{aa} 
\bibliography{main} 

\end{document}